# Harnessing the VO$_2$ Phase Transition for Automatic Gain Control in Transimpedance Amplifiers

Amir Gildor[1], https://orcid.org/0009-0005-5847-6612; Sariel Hodisan[2], https://orcid.org/0009-0005-3758-8981;

Shahar Kvatinsky[2], https://orcid.org/0000-0001-7277-7271; Yoav Kalcheim[1], https://orcid.org/0000-0002-1489-0505

[1] *Department of Materials Science and Engineering, Technion - Israel Institute of Technology,* Haifa, Israel

[2] *Viterbi Faculty of Electrical and Computer Engineering, Technion - Israel Institute of Technology,* Haifa, Israel

*Abstract*—Transimpedance amplifiers (TIAs) are essential in sensor electronics, converting input currents into output voltages. Conventional TIAs utilize fixed-gain resistors, which saturate under high input currents and consequently result in undesirable recovery times. To overcome this limitation, volatile resistive switching devices have emerged as a promising alternative, offering intrinsic automatic gain control (AGC). Among these, vanadium dioxide (VO$_2$) devices stand out for their reversible insulator–metal transition (IMT), producing abrupt, energy-efficient resistance changes near the transition temperature ($T_C \approx 67$ °C). In this work, a switching device was fabricated by sputtering a VO$_2$ thin film and patterning ~200 nm electrode gaps atop it. Before integrating this device into the TIA circuit, its switching dynamics were characterized under electrical pulse excitation. Slightly exceeding the temperature-dependent IMT threshold voltage ($V_{th}$) yielded fast and reproducible switching. Complementary pump–probe measurements showed that operating well below $T_C$ effectively suppresses short-term memory effects linked to the stochastic nature of the first-order transition. Leveraging these insights, a custom VO$_2$-based TIA was developed, demonstrating variable gain and AGC functionality. Furthermore, applying a constant DC current bias during switching induced self-sustained oscillations (~2 pJ per spike) with frequencies up to ~60 MHz, consistent with the thermal timescale of the VO$_2$ devices. Overall, these results provide a detailed understanding of VO$_2$ switching dynamics and demonstrate their potential for enabling compact, energy-efficient AGC in high-speed TIAs for advanced sensing applications.

*Keywords—Transimpedance amplifier (TIA), automatic gain control (AGC), insulator-metal transition (IMT), volatile resistive switching, vanadium dioxide (VO$_2$).*

## I. Introduction

Transimpedance amplifiers (TIAs) serve as critical analog interface circuits for the conversion of input currents—typically generated by photodetectors, biosensors, and other current-output devices— into voltage signals that can be amplified, digitized, and processed in electronic systems [1]. The core function of a TIA relies on a low input impedance configuration, often realized using an operational amplifier with a feedback loop incorporating a gain resistor. In this configuration, the input current is directed into the inverting terminal of the operational amplifier, where it is converted into a voltage at the output by the feedback resistor, according to Ohm's law ($V_{out} = -I_{in} \times R_f$). This fundamental structure facilitates amplification of weak currents while maintaining a virtually zero voltage at the input node, guaranteeing minimal loading of the signal source [2]. The feedback resistor's value, $R_f$, directly sets the amplifier's transimpedance gain, but also influences the noise performance and bandwidth [3], imposing a trade-off between high sensitivity and high-speed operation [4].

The high dynamic range often required in modern applications such as optical receivers [5], single-photon detection [6], and high-throughput biosensing [7] greatly complicates TIA design. In these applications, the TIA must maintain low noise and high gain for small signals, yet avoid saturation, distortion, or "blind" periods in the presence of large input currents, common under conditions of strong illumination or rapid signal bursts [2]. Saturation occurs when the input current exceeds the designed dynamic range, causing the operational amplifier to output voltages at or near its supply rails, pausing further signal acquisition until recovery [8]. This scenario is particularly problematic in state-of-the-art detectors designed for optical communication [9], quantum technologies [10], LiDAR [11], high-speed imaging [12], or neuroscience [13], all of which demand both sensitivity and broad dynamic range for accurate signal representation.

Various strategies have been proposed to extend the dynamic range of TIAs. Traditional approaches employ multiple gain stages [14], [15] or digitally switched feedback resistors [16] to accommodate different input signal levels. However, such designs increase system complexity, switching latency and often limits the achievable bandwidth. Another frequently used method involves the inclusion of active devices, such as field-effect transistors (FETs), in parallel or in the feedback path, offering adjustable gain through gate voltage modulation [17]. While providing dynamic tuning, these methods can introduce additional noise, power consumption, and often demand complex calibration and control circuits.

More contemporary solutions focus on implementing automatic gain control (AGC), in which the effective transimpedance gain adapts in real-time to the input current level [18], [19]. Nonvolatile resistive switching elements, such as valence-change mechanism (VCM) memristors [20], have been explored as dynamic gain elements for TIAs, exploiting their variable resistance properties to achieve programmable, even self-adaptive, gain [21]. However, such AGC circuits employing non-volatile switching require external reset circuitry for operation in repetitive or high-rate signal environments, thereby adding integration complexity and limiting operation times.

We suggest to use an alternative AGC strategy using volatile resistive switching devices based on an insulator-metal transition (IMT). To this end we employ thin film devices based on vanadium dioxide ($VO_2$) - a widely studied material within the Mott insulators family [22]. Nano-scale $VO_2$-based switching devices have been investigated in diverse contexts, such as memory elements [23], oscillatory units [24], [25], [26], [27], [28], [29], artificial neurons [30] and signal processing circuits [31], [32] - all exploiting the IMT properties.

$VO_2$ undergoes a first-order IMT at a transition temperature ($T_C$) of ~67 °C [33], close to ambient temperature. This transition is marked by an abrupt drop in electrical resistance by several orders of magnitude, involving a structural shift from a low-temperature monoclinic insulating phase to a high-temperature rutile metallic phase [22], [34], [35]. The transition does not proceed uniformly; instead, it occurs through an abrupt transitions of nanoscale domains between insulating and metallic states [36]. The IMT can be triggered thermally via uniform heating or locally through Joule heating induced by current injection or applied voltage, forming conductive filaments (Figure 1.a) [36], [37], [38]. Notably, the spatial distribution and dynamics of these domains are inherently stochastic, influenced by thermal fluctuations, local strain, and microstructural variations in the film [39], [40], [41]. Another key feature of this transition is its volatility— upon removal of the stimulus, the material rapidly relaxes to the insulating state [42].

Thus, $VO_2$-based switching devices offer several intrinsic advantages in adaptive electronic systems, particularly for AGC in TIAs: the resistance change across the IMT can span over several orders of magnitude [43], allowing the TIA to accommodate a wide dynamic range of input currents; the volatile nature of the phase transition facilitates automatic recovery to the high-resistance state [30] (automatic gain resetting), supporting repetitive and cyclical operation for high-speed detection, without the need for external reset or refresh mechanisms; $VO_2$ devices also exhibit remarkable endurance, sustaining billions of phase transition cycles without appreciable performance degradation [43]; and the speed of the IMT in $VO_2$ is on the order of tens of nanoseconds or faster [44], supporting rapid response. Therefore, understanding the rapid dynamics of $VO_2$ switching devices is necessary to the development of a robust, energy-efficient $VO_2$-based TIA for high-speed applications.

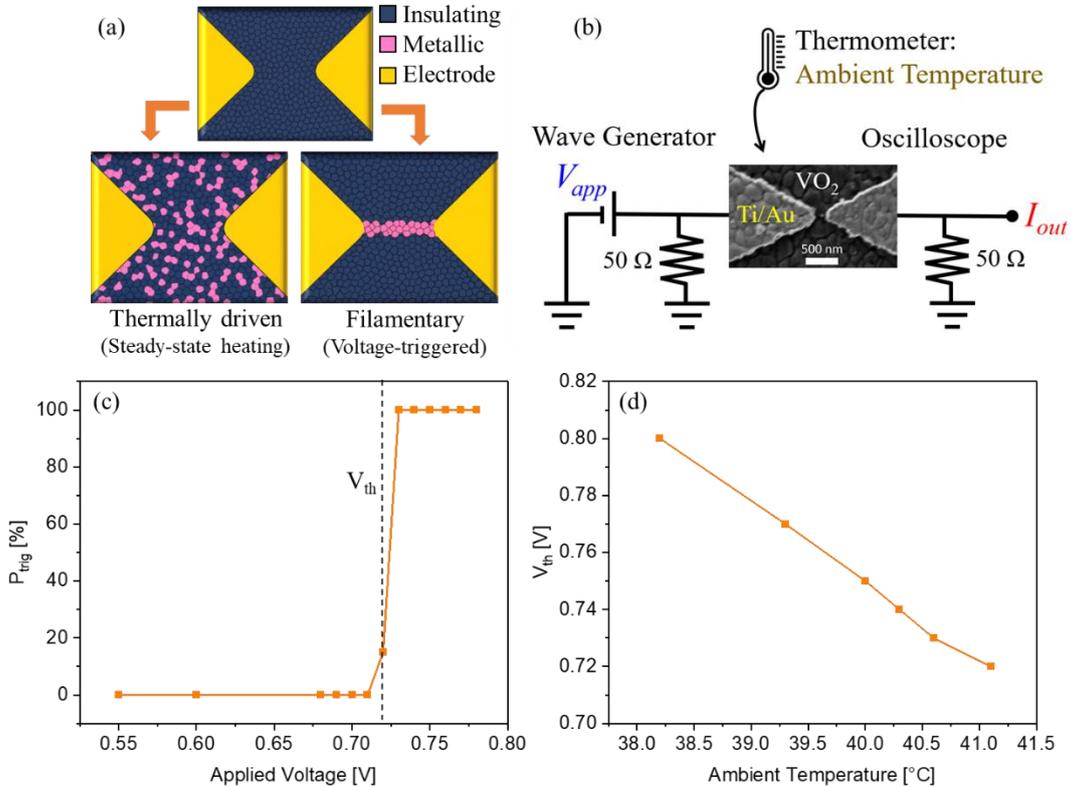

*Figure 1: **Switching threshold voltage ($V_{th}$) measurements.** (a) Illustration of the switching mechanisms contributing to the phase transition of $VO_2$ devices: thermally-driven formation of metallic domains under uniform heating or filament formation under an applied electric field between the electrodes. Both mechanisms may occur simultaneously, lowering $V_{th}$ closer to the*

*transition temperature by lowering the energy barrier to form metallic domains. (b) Schematic of the measurement setup, where a function generator applied voltage pulses across a VO₂ switching device, while an oscilloscope recorded the output current. Additionally, the ambient temperature was recorded by a thermometer. (c) The triggering probability ($P_{trig}$) transitioned sharply from 0 to 100%, when the applied voltage slightly surpassed $V_{th}$. (d) dependence of $V_{th}$ on ambient temperature.*

## II. RESULTS AND DISCUSSION

A vanadium dioxide (VO₂) switching device was fabricated by sputtering a thin film (~150 nm) of the oxide on a (11-20)-oriented (A-cut) sapphire substrate and forming two metallic electrodes atop it. The film had a highly-oriented crystalline structure, where monoclinic (100) planes were preferred, as confirmed by a 2theta-omega X-ray diffraction (XRD) measurement (Figure 2). Standard titanium-gold (200 Å/700 Å) electrodes, separated by a ~200 nm gap (Figure 1.b), were patterned on top of the film by electron beam lithography (EBL) and metal evaporation. This electrode composition is characterized by optimal adhesion, electronic conductance and durability properties [45], [46]. To characterize the dynamics of this switching device, its electrodes were initially soldered to a custom-made printed circuit board (PCB) designed for high-speed voltage-current measurements. This measurement platform was equipped with a current compliance mechanism, responding in nanoseconds (see Methods section) [47], to control current at fast timescales and prevent device burnout at sudden resistance drops. The PCB was connected to a function generator to supply the input voltage pulses, and was routed to an oscilloscope which measured the voltage applied ($V_{app}$) on the device and the output current ($I_{out}$), with a temporal resolution of 0.5 ns. Additionally, the PCB was placed in a closed enclosure equipped with a thermometer next to the device, to monitor the ambient temperature during operation (Figure 1.b).

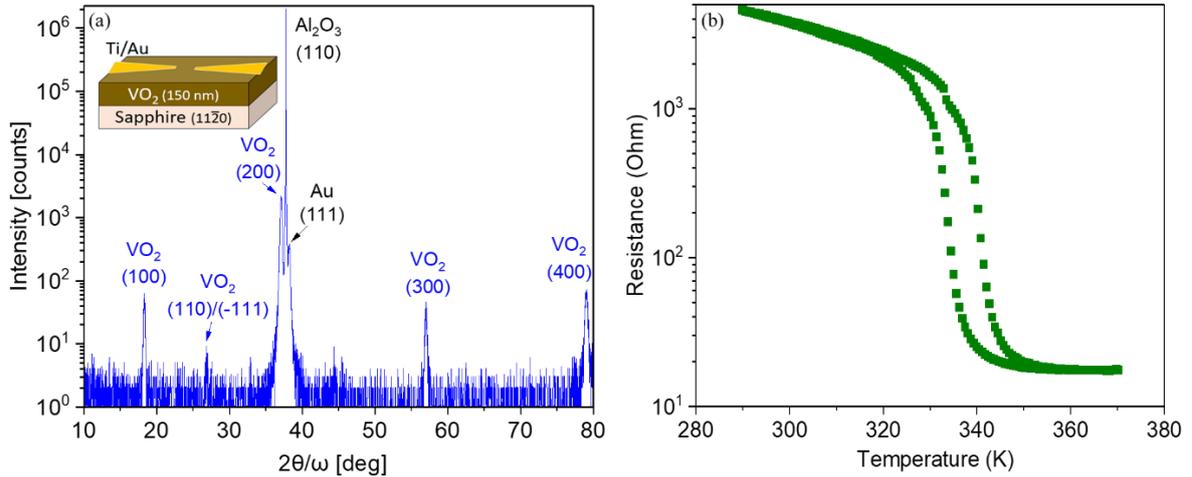

*Figure 2: X-ray diffraction (XRD) and Resistance-Temperature (R-T) Characterization of a VO₂ device. (a) The diffraction pattern of the devices indicates the highly-oriented crystalline structure of the VO₂ film on (11-20)-oriented (A-cut) sapphire substrate, with a strong preference for the (100) orientation. Inset: Schematic sideview of the fabricated device. (b) A steady-state R-T curve, identifying $T_C$ at ~340 K.*

First, the triggering probability of the device ($P_{trig}$) was found at different $V_{app}$ values, ranging between 0.55-0.80 V. For each voltage amplitude, 200 square voltage pulses with a pulse width of 800 ns each were applied on the device while recording $V_{app}$ and $I_{out}$ with respect to time (Figure 3.a). This was done with a delay time ($\tau$) of 1 ms between pulses and at an ambient temperature of 41.1 °C, far below $T_C$, to minimize possible memory effects [48]. $P_{trig}$ was determined for each $V_{app}$ as the percentage of triggering events in $I_{out}$, allowing the identification of the switching threshold voltage ($V_{th}$). Switching is consistently triggered above this threshold, exhibiting a sharp transition of $P_{trig}$ from zero to one hundred percent (Figure 1.c). The width of this transition is 20 mV, representing less than 3% of $V_{th}$. Additionally, $V_{th}$ decreased by ~3.5% per degree with the rise of ambient temperature near the device (Figure 1.d). These findings can be explained by the lower energy barrier for switching when approaching the IMT temperature, demonstrating the importance of accurate temperature control to adjust devices' switching properties. While this might be challenging in traditional integrated circuits, novel designs may exploit it as another "tuning knob" of the switching dynamics.

To investigate the temporal dynamics of these switching devices, we measured the two characteristic time intervals of a triggering pulse initiation (Figure 3.a). The first is the incubation time ($t_{inc}$), defined as the delay between the start of the voltage pulse and the onset of current rise. The second is the Rise time ($t_{rise}$), representing the duration from the onset of switching until current has stabilized.

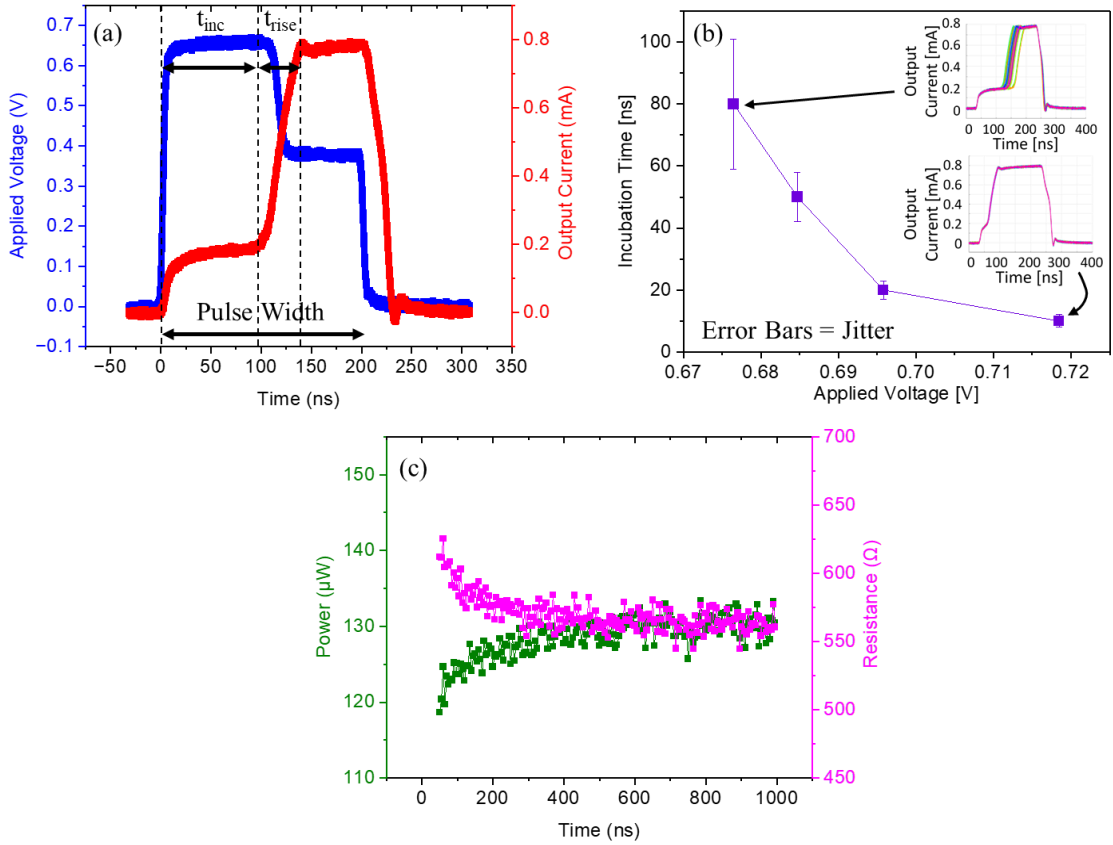

*Figure 3: Characteristic temporal dynamics of current-limited switching upon the application of voltage pulses. (a) A representative output current pulse and the definitions of incubation time ($t_{inc}$), rise time ($t_{rise}$) and the programmable pulse width. Current amplitude was limited to 0.8 mA by a current compliance mechanism. (b) Shorter incubation times were measured with the application of higher voltages. Error bars indicate the spread of $t_{inc}$ ("jitter"), which similarly diminishes with voltage rise. Insets: overlays of 200 output current pulses, demonstrating the faster, more stable incubation time as voltage increases. (c) Comparison of power and resistance during a 1000 ns pulse. Resistance creeps down while power creep up within the first 300 ns, until steady state is established.*

The average $t_{inc}$ was extracted from $I_{out}$ traces following 200 consecutive voltage pulses applied with amplitudes ($V_{app}$) exceeding a $V_{th}$ of 0.67 V, each with a pulse width of 200 ns and limited to 0.8 mA by the current compliance mechanism (Figure 3.b). The jitter of $I_{out}$ around the mean value was also quantified to characterize the statistical variation of $t_{inc}$ as a function of $V_{app}$. Near $V_{th}$, $t_{inc}$ exhibits significant jitter, but it stabilizes to approximately 10 ns as $V_{app}$ increases. Notably, this stabilization occurs with less than a 10% increase above $V_{th}$, indicating that rapid and reproducible switching can be achieved by only slightly exceeding the threshold, despite its temperature dependence.

In contrast to $t_{inc}$, $t_{rise}$ remains approximately constant at 30 ns, where the system's limitation is approximately 5 ns due to the finite analog bandwidth of the on-board measurement amplifiers. This suggests that while the initiation of the transition is voltage-dependent, the subsequent rise-time may be governed by the intrinsic kinetics of filament evolution rather than measurement circuitry alone. Specifically, once the IMT is triggered, the expansion and stabilization of the metallic filament are likely dominated by the thermal diffusivity of the $VO_2$ layer and the surrounding substrate. The observed constancy of $t_{rise}$ across varying $V_{app}$ values indicates that this lateral filament growth rate is independent of the voltage triggering, probably due to the current limiting circuitry. Together, $t_{inc}+t_{rise}$ define the activation timescale of the device and demonstrate its capability for short-pulse operation.

We note that the timescales required to reach steady state exceed the thermal timescales of the device by nearly an order of magnitude, pointing towards the thermally activated nature of switching which plays an important role in both the filament formation and relaxation processes. During switching from the high to low resistance state, a filament is formed and a new temperature distribution is established within a few tens of nanoseconds. The filament is maintained above $T_C$ by Joule heating while the surroundings insulating domains are slightly cooler. However, even after thermalization is reached, filament-adjacent insulating domains may switch into the metallic state by overcoming the free energy barrier in a thermally activated process. This barrier is a direct manifestation of the first order nature of the IMT in VO$_2$. The resulting widening of the filament manifests as a gradual decrease in resistance until ~300 ns (Figure 3.c). Assuming an expansion where the filament length and resistivity remain constant, the change in resistance scales inversely with the cross-sectional area ($A$) according to $R = \rho * \frac{l}{A}$. Given the measured ratio of the final to initial resistance in the low-resistance state (Figure 3.c), $\frac{R_{final}}{R_{initial}} = \frac{A_{initial}}{A_{final}}$, these results indicate a total increase in the filament's cross-sectional area of approximately 10%. We note that slow filament widening was previously imaged [49] but occurred over two orders-of-magnitude longer timescales, probably due to the use of a fixed voltage source throughout the pulse rather than the present current limiting circuitry.

In a TIA, the VO$_2$ device may be subjected to multiple switching events separated by short timescales. If the device does not fully relax to the insulating state after a switching event, a subsequent trigger may induce unwanted switching even if the voltage does not reach $V_{th}$. As shown by del Valle *et al.* [48] isolated metallic domains may persist in the device after switching with little effect on its resistance, but they may substantially decrease the switching voltage since the isolated metallic domains act as hot-spots which concentrate current and field thereby facilitating the IMT. This effect becomes more pronounced as the temperature approaches the metal-to-insulator transition (MIT) temperature ($T_{MIT}$) due to the weaker restoring force into the insulating state which renders the relaxation times larger. To elucidate relaxation dynamics and their possible effect on switching voltage, we examined the possibility of subthreshold switching due to past switching events in our devices by all-electrical pump-probe measurements. In these experiments, an initial super-threshold voltage pulse (pump) was applied to trigger switching followed by a subthreshold pulse (probe) to detect triggering due to incomplete relaxation after a certain delay time τ (Figure 4). The super-threshold pulse was 1.12 $V_{th}$ to ensure switching by the pump while the probe was set to 0.99 $V_{th}$ to maximize the detection of incomplete relaxation in the device. A waiting time of 1 ms was set between each pump-probe couple ensure a complete relaxation before every pump pulse. Three pump-probe delay times (τ=100, 200, 500 ns) were examined and 200 repetitions were performed for each delay time. As a control, the results were compared to the probability of switching from isolated probe pulses (Table 1). We found that for τ of up to 200 ns, probe pulses exhibited very low triggering probabilities ($P_{trig}$< 1.5%), whereas for τ = 500 ns the probability dropped to 0%. When compared to previous results [48] which showed relaxation times up to milliseconds, these results show orders of magnitude faster relaxation. This is attributed to the lower temperature which was used in this experiment compared to previous work, which results in enhanced cooling rate and driving force to the insulating phase.

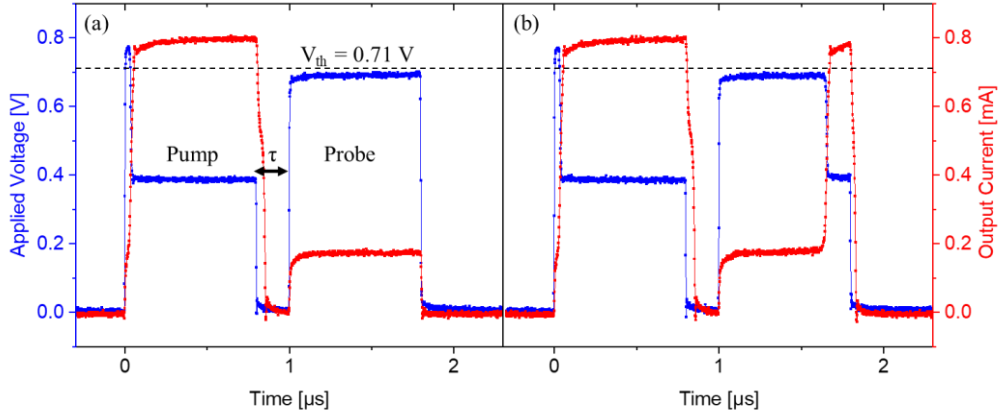

*Figure 4: **Investigation of relaxation dynamics via double pulse pump-probe measurements.** The applied voltage (blue) shows two pulses: the first is set to a voltage exceeding $V_{th}$ to ensure it always induces switching, as shown by the large rise in output current (red). The drop in applied voltage during the pulse is due to the current compliance. After the initial pulse ends the device is held for a time τ before a second subthreshold pulse is applied. Subthreshold switching in the second pulse indicates incomplete relaxation back to the insulating phase after switching in the first pulse. Out of 200 cycles tested with a subthreshold voltage of 0.99$V_{th}$, (a) represents the common case with no subthreshold switching, while (b) displays one of the scarce (only 1.5%) subthreshold switching events. The measurement temperature was 41.4 °C.*

*Table 1: **Pump-probe measurements at different delay times (τ)**. A total of 200 cycles were conducted, where minor memory effects (1.5 percent triggering at the probe pulse) were only observed at τ=100 ns and τ=200 ns. At τ=500 ns and above, no memory effects were observed.*

| Single Pulse (Control) | Pump-Probe Measurements | | |
|---|---|---|---|
| | *τ = 100 ns* | *τ = 200 ns* | *τ ≥ 500 ns* |
| No memory effects ($P_{trig}$=0%) | Minor memory effects ($P_{trig}$=1.5%) | Minor memory effects ($P_{trig}$=1.5%) | No memory effects ($P_{trig}$=0%) |

After characterizing the switching dynamics of the device under voltage pulses, it was mounted on another custom-made PCB to be tested within the TIA circuitry, replacing the constant gain resistor (Figure 6.a, inset). This apparatus receives a voltage pulse from a function generator, converts it into a current pulse, and applies it to the VO$_2$-based device, while the voltage output is recorded by an oscilloscope. In a standard TIA, the input current develops a voltage across the feedback resistor $R_f$, which sets the gain ($V_{out} = -I_{in}*R_f$). By replacing $R_f$ with VO$_2$, the feedback element becomes nonlinear and current-dependent. At low input signals, the VO$_2$ is in the non-perturbed high resistance state, producing a high output voltage. As the input signal rises, Joule heating lowers the VO$_2$ resistance, reducing the voltage drop across it and hence the amplifier's gain. When the input is sufficient to trigger the IMT, VO$_2$ abruptly switches to a low-resistance state, clamping the feedback path and sharply decreasing the output voltage. This provides an AGC mechanism that prevents TIA saturation.

The bandwidth of the VO$_2$-based TIA was first evaluated by applying sine waves, sweeping the frequency from 1 MHz to 1 GHz using a network analyzer. Measurements were repeated at increasing input powers, starting from -36 dBm and up to 0 dBm, to determine the minimum power required to trigger the IMT (Figure 5.a). As the input power increased, heating of the device gradually reduced its resistance – observed as a decrease in gain – until switching occurs at about -14 dBm. To prevent excessive currents during the transition, a current-limiting mechanism was set to 1 mA, a value which facilitates large gain variations upon switching while preventing damage to the device. Notably, above −14 dBm a frequency-dependent drop in gain emerges, with the minimum shifting from ~10 MHz to ~15 MHz as the input power increases (see dashed line in Figure 5). This behavior likely originates from synchronization between the period of the external drive and the intrinsic thermal timescale of the device. In the limit where the drive period is much longer than the thermal timescale (low frequency), the device resistance closely follows the power oscillations, which occur at twice the drive frequency (since $P \propto V^2$). Under these conditions, the device dissipates sufficient heat during each cycle to return to the high-resistance state between oscillations. As the drive period approaches the thermal timescale, the cooling time becomes comparable to the excitation period, preventing the device from fully recovering between cycles. This incomplete recovery leads to a reduction in the average gain. At higher frequencies, where the drive period is shorter than the thermal timescale, the device effectively averages over the power oscillations. This reduces the efficiency of driving the system into the low-resistance state, resulting in a recovery of higher gain. The shift of the gain dip toward higher frequencies with increasing input power, is consistent with a smaller thermal cycle – the temperature variation diminishes as the low resistance state is approached. A similar trend is also observed under constant current driving as discussed below.

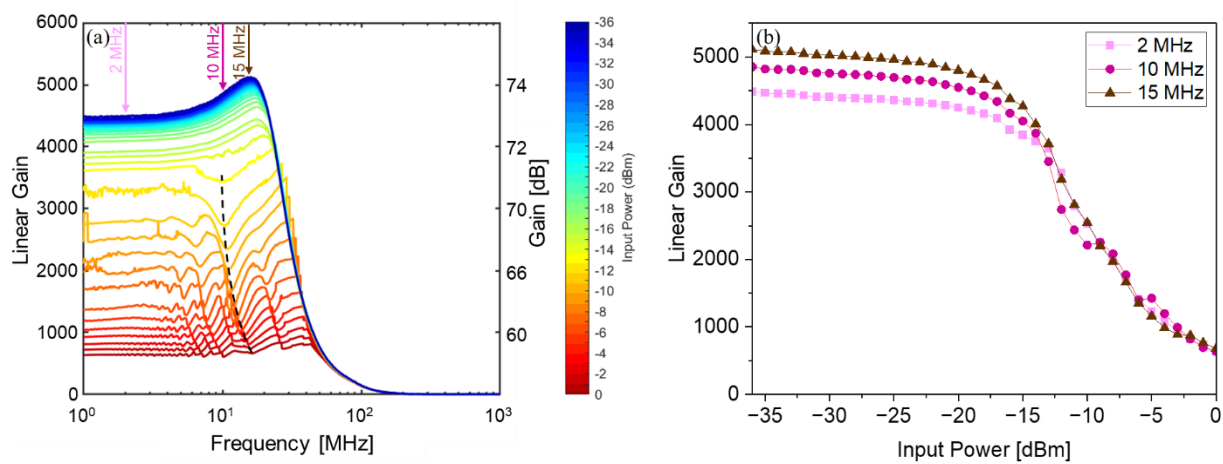

*Figure 5: **Bandwidth in VO$_2$-based TIA - frequency response and dynamic range.** (a) TIA gain as a function of frequency (1 MHz to 1 GHz) for various input power levels (left: linear scale; right: dB scale). With increasing input power Joule heating lowers the device resistance and gain until the IMT is triggered above -14 dBm. Current compliance is maintained at 1 mA to ensure circuit stability. Gain plotted on a linear scale, highlighting a drop at ~10 MHz, which correlates with the characteristic switching oscillation frequency (Figure 7.d). (b) Gain as a function of power at fixed frequencies, indicating a sharp gain drop above -14 dBm.*

To study the response of the VO$_2$-based TIA to current-driving, we used square voltage pulses of varying amplitudes from a function generator, which were locally converted into constant and stable current pulses on the PCB using a LT1228 transconductance amplifier. The resulting output voltage on the device was recorded with an oscilloscope. Input currents ranged from ~40 µA to ~940 µA (i.e., up until the current limit of 1 mA), with a pulse width of 300 ns each (Figure 6). Offset currents measured without an input signal reached only 40 µA, yielding a negligible output baseline of up to ~10 mV.

At low input currents, the device behaves as a standard resistor, where the output voltage rises during the pulse and returns to baseline upon its conclusion (Figure 6.a). In contrast, at the highest input currents applied, the $VO_2$ undergoes a complete IMT, resulting in a minimal voltage rise as the device remains in the low-resistance state (Figure 6.c). At intermediate current levels, constant pulses induce rapid autonomous oscillations between the insulating and metallic states, generating voltage oscillations on a timescale of tens of nanoseconds (Figure 6.b).

This oscillatory behavior arises from self-sustained electrothermal cycles and is promising for applications in oscillatory neural networks (ONNs) [27], [50]. When the device is in its high-resistance state, the application of a DC current induces Joule heating ($I^2R$), raising the temperature until the IMT is triggered. Upon transitioning to the low-resistance state, the $I^2R$ heating is significantly reduced, allowing the accumulated heat to dissipate into the surroundings. This cooling leads to a MIT, reverting the device to its high-resistance state and allowing the cycle to repeat as long as the current is maintained [50], [51]. It is important to note that this autonomous behavior, occurring *without* external capacitors or resistors, is exclusive to the current-driven mode [51]. The precise current stabilization provided by our TIA circuitry was instrumental in enabling the measurement of these high-speed patterns, which would otherwise be obscured by voltage-driven instabilities.

The performance of this $VO_2$-based oscillator is compared with state-of-the-art CMOS oscillators and previously reported $VO_2$ switching devices in Table 2. The measured frequency of ~ 60 MHz is consistent with the thermal response timescales ($t_{inc}$, $t_{rise}$) characterized earlier. This frequency exceeds reported $VO_2$ oscillators by one to four orders of magnitude [52], [53], [50] and approaches the performance of current-controlled CMOS oscillators [54]. While voltage-controlled CMOS architectures [55] achieve higher frequencies (up to 30 GHz), they require significantly higher power (1500 µW) in comparison to our devices (60-100 µW in the oscillatory window, Figure 7.c). Additionally, voltage-mode CMOS devices are more complicated to integrate into current-mode TIAs. In contrast, our $VO_2$ device demonstrates a low energy consumption of approximately 1.5 pJ per oscillation - calculated as the area under the power oscillation ($P = V_{out} \cdot I_{in}$) (Figure 7.a, Figure 7.b). This represents a near two-order-of-magnitude improvement in efficiency over earlier $VO_2$ implementations [52], [53], [50], positioning this technology as a compelling candidate for low-power, current-driven hardware.

The ability to resolve these high-speed oscillatory patterns was enabled by the precise current stabilization of this TIA circuitry [21], [47], originally designed for signal amplification. By preventing the voltage-driven instabilities common in standalone devices, the current stabilization stage provides a robust platform for both TIA operation and the characterization of fast volatile dynamics. While the immediate functional value of this integration lies in the realization of a compact AGC mechanism, the low-energy oscillations and autonomous switching can be highly attractive for other hardware applications and oscillatory units, such as neuromorphic devices [26], [27], [30].

Finally, these results demonstrate that the $VO_2$ element acts as an active, self-regulating feedback component within the TIA. By transitioning to a low-resistance state at high input current, the $VO_2$ device inherently reduces the amplifier gain, thereby preventing output saturation and extending the dynamic range without the need for complex, external feedback loops or multi-stage CMOS control circuitry. This synergy between Mott physics and analog circuitry provides a design paradigm for the next generation of adaptive, high-speed front-end electronics.

*Table 2: **Performance comparison of $VO_2$ oscillators and CMOS technologies.** Operating frequencies, peak power, and energy consumption per oscillation event are summarized. In this work, $VO_2$-based device demonstrates a significant increase in frequency compared to prior $VO_2$ benchmarks while simultaneously reducing energy consumption to ~ 1.5 pJ per spike, moving closer to the efficiency and speed of specialized current-controlled CMOS architectures.*

| Technology | Operation Frequency [MHz] | Power Peak [µW] | Energy/Oscillation [pJ] |
|---|---|---|---|
| Voltage-controlled CMOS [55] | 30000 | 1500 | 0.83 |
| Current-controlled CMOS [54] | 80 | 60 | 0.11 |
| $VO_2$ [52] | 0.0003 | 0.16 | 100 |
| $VO_2$ [53] | 0.1 | ~23 | ~110 |
| $VO_2$ [50] | 2.1 | ~370 | ~80 |
| $VO_2$ - This work | **~40-60** | **60-100** | **~1.2-1.9** |

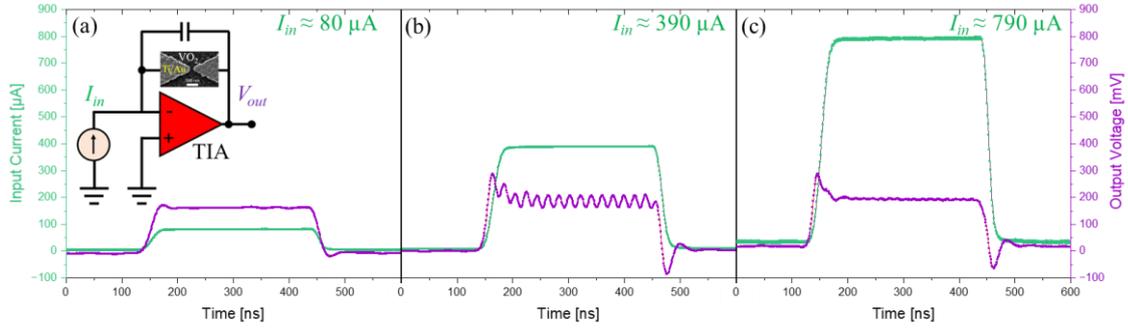

*Figure 6: **Oscillatory output pattern of a $VO_2$-based TIA under DC current input.** (a)-(c) DC current pulses with different amplitudes were applied, by the circuit illustrated in the inset, utilizing a $VO_2$-based switching device as an automatic gain control (AGC). (a) Low currents did not trigger the IMT. (b) Beyond a threshold current of ~200 µA (Figure 7.c), oscillatory output patterns appeared, with a period of ~20 ns, similar to the thermal timescale. (c) Above ~600 µA (Figure 7.d), the oscillations were below the noise limit, demonstrating stabilization of the low-resistance state.*

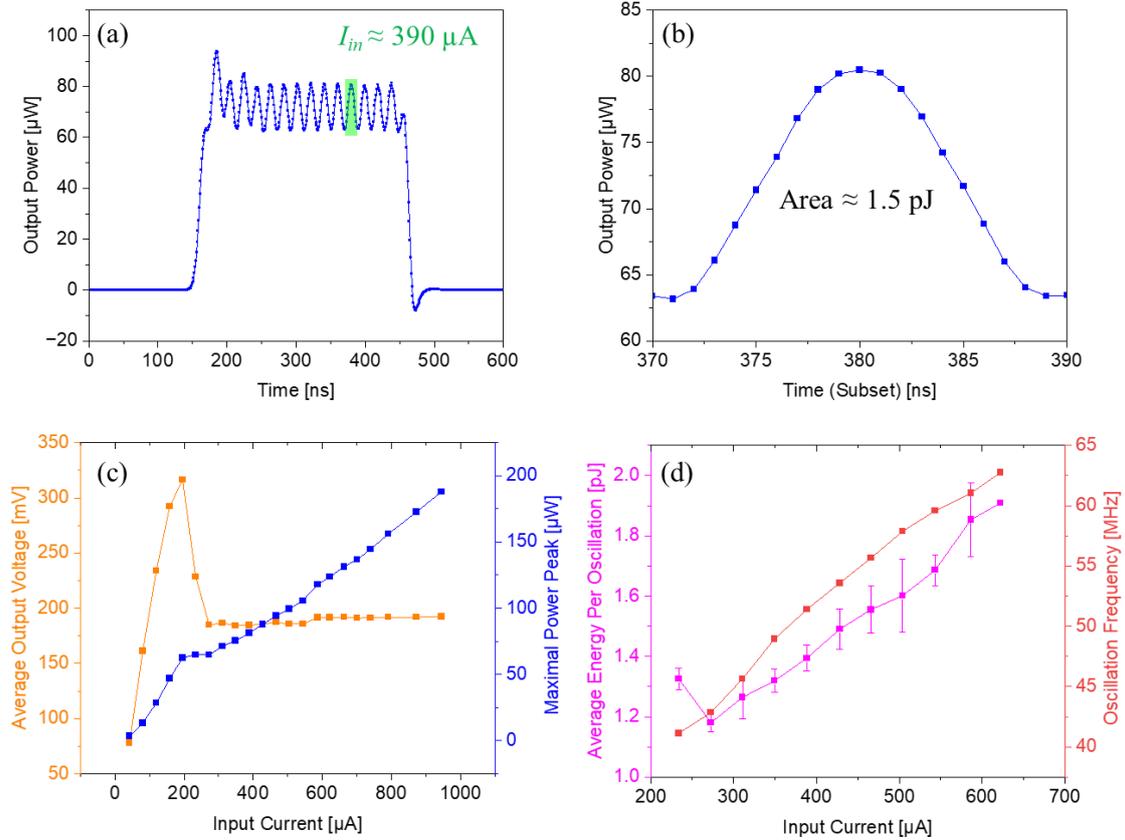

*Figure 7: **Energy and operational window of VO₂ oscillatory dynamics.** (a) Time-resolved output power ($P=V_{out}*I_{in}$) derived from the voltage spikes in Figure 6.b. The peak marked by the green rectangle is viewed in detail in (b), where the area under the curve corresponds to an energy consumption of ~1.5 pJ per oscillation. (c) Average output voltage vs. input current, identifying the IMT threshold at ~200 µA, where the oscillatory behavior begins. (d) Evolution of the average energy per oscillation and the oscillation frequency as a function of input current, within the range of oscillatory behavior (~200–600 µA).*

## III. Conclusion

A VO$_2$-based resistive switching device was fabricated, characterized, and integrated into a transimpedance amplifier to provide automatic gain control with inherent self-recovery. Unlike non-volatile resistive switches that require complex additional reset circuitry, the volatile nature of this device enables automatic gain resetting, a critical feature for continuous high-speed signal acquisition. With combined incubation and rise times of ~40 ns, the device supports short-pulse operation. Short-term memory effects and stochasticity were identified through relaxation dynamics, but do not compromise device speed. By employing a strict current limit to eliminate power overshoot and creep, this work reveals dynamic effects beyond standard thermal timescales, specifically observing filament widening and MIT relaxation effects. Furthermore, this configuration facilitated the generation of extremely fast and energy-efficient oscillations under DC excitation, reaching frequencies of ~60 MHz and energy consumption of ~1.5 pJ per event, comparable to state-of-the-art oscillatory devices. These results demonstrate that VO$_2$ switching devices offer a robust solution for extending dynamic range in high-speed amplifiers and oscillatory devices.

## IV. Methods

### A. Device Fabrication and Characterization

A ~150 nm VO$_2$ thin film was deposited on a (110)-oriented sapphire substrate (A-cut) by reactive RF magnetron sputtering (AJA International Inc., USA) using a sintered V$_2$O$_3$ target at an oxygen-argon atmosphere. Its crystalline structure was characterized with a 2θ/ω measurement by a high-resolution X-ray diffractometer (XRD; Smart-Lab 9 kW, Rigaku, Japan), using a rotating anode X-ray source of 1.54Å (Cu Kα) wavelength and a Ge-2x200 monochromator, installed between the source and the sample. After fabricating and characterizing the film, two Ti/Au electrodes (200/700 Å) separated by a ~200 nm gap were patterned on top of it using a standard electron beam lithography (EBL) process, by an EBPG 5200 EBL system (Raith, Germany) and a BAK-501A E-beam Evaporator (Evatech, Switzerland) for metal deposition. Finally, a structural characterization was carried out by an Ultra-Plus FEG-SEM (Zeiss, Germany) for High-Resolution Scanning Electron Microscopy (HRSEM, Figure 1.b). A schematic illustration of the fabricated device is shown in Figure 2.

*B. High-Speed Resistive Switching Measurements*

An advanced measurement platform, equipped with an extremely rapid current compliance mechanism, was used to perform high-speed voltage-current (V-I) measurements. This setup included a function generator (AFG31000, Tektronix, USA), an oscilloscope (6 Series B MSO MSO64B, Tektronix, USA) and a custom-made printed circuit board (PCB) on which the $VO_2$-based device was placed. The compliance mechanism was implemented by placing the switching device in series with a gigahertz-speed common-base (CB) stage implemented using an NPN transistor, as demonstrated by Hennen *et al.* [47]. The bias current of the CB stage was adjustable and defined the compliance current. Because the CB stage functioned as a current buffer, the current through the switching device was also forced through this stage. When the current through the device approached or exceeded the CB bias current, the CB stage entered cutoff within nanoseconds due to its gigahertz bandwidth. Consequently, the collector voltage of the CB stage rose sharply. Since this collector node was directly connected to the second electrode of the $VO_2$ device, the voltage drop across the device was abruptly reduced. This rapid reduction in voltage limited the device current and thus protected it from damage with a nanosecond-scale response time.

For the TIA operation, a second custom-made PCB was used, with the same function generator and oscilloscope. A CB stage was placed before the TIA to provide protection when the input current reached approximately 1 mA. Since the CB stage is not ideal, part of its bias current was diverted into the TIA rather than flowing entirely through the CB stage. This leakage accounts for the offset current observed even in the absence of an applied input signal, but it was minimized to a negligible value of up to 40 µA. Once current compliance was activated on the TIA board, the calculated power values in dBm became invalid. Beyond this point, further increases in the input signal did not result in higher TIA output, as the compliance circuit clipped the current before it reached the TIA. In this setup, a network analyzer (SVA1032X, Siglent Technologies, USA) was used for bandwidth measurements (Figure 5).

ACKNOWLEDGMENT

Devices were fabricated with the support of the Technion's Micro-Nano Fabrication Unit (MNFU). The authors wish to thank Mr. Ilan Lipschutz, ASIC[2] lab manager, for his guidance in operating the measurements. Devices' characterization was done at the Technion's Electron Microscopy Center (EMC), Faculty of Materials Science and Engineering.

This work has been funded by the European Union (ERC, MOTTSWITCH, 101039986) and Israel Innovation Authority KAMIN project no. 81755. Views and opinions expressed are however those of the author(s) only and do not necessarily reflect those of the European Union or the European Research Council Executive Agency. Neither the European Union nor the granting authority can be held responsible for them.